\documentclass[letterpaper, 10 pt, conference]{ieeeconf}
\IEEEoverridecommandlockouts
\usepackage{graphicx}          
\usepackage{subfig}
\graphicspath{{figures/}}

\usepackage{ifthen} 
\newboolean{BGRAPHPDF}
\setboolean{BGRAPHPDF}{true}
\newboolean{TECHREP}
\setboolean{TECHREP}{true}

\ifthenelse {\boolean{BGRAPHPDF}}
{
\usepackage[ruled]{algorithm}
\usepackage{algpseudocode}
}
{
\usepackage{vaucanson-g}
}

\usepackage{cite}     
\usepackage{fainekos-macros}

\title{\LARGE \bf
Revision of Specification Automata under Quantitative Preferences
}

\author{Kangjin Kim and Georgios Fainekos%
\thanks{This work has been partially supported by award NSF CNS 1116136.}%
\thanks{K. Kim and G. Fainekos are with the School of Computing, Informatics and Decision Systems Engineering, Arizona State University, Tempe, AZ 85281, USA { \tt\small \{Kangjin.Kim,fainekos\}@asu.edu} }%
}

\begin{document}

\maketitle
\thispagestyle{empty}
\pagestyle{empty}


\begin{abstract} 
We study the problem of revising specifications with preferences for automata based control synthesis problems.
In this class of revision problems, the user provides a numerical ranking of the desirability of the subgoals in their specifications.
When the specification cannot be satisfied on the system, then our algorithms automatically revise the specification so that the least desirable user goals are removed from the specification.
We propose two different versions of the revision problem with preferences.
In the first version, the algorithm returns an exact solution while in the second version the algorithm is an approximation algorithm with non-constant approximation ratio.
Finally, we demonstrate the scalability of our algorithms and we experimentally study the approximation ratio of the approximation algorithm on random problem instances.
\end{abstract}



\section{Introduction}

Linear Temporal Logic (LTL) has been widely adopted as a high-level specification language for robotic behaviors (see \cite{KressGazit11ram} for a recent overview).
The wide spread adoption of LTL can be attributed to the tractable algorithms that can solve automation problems related to robotics (see \cite{KressGazit11ram}) and the connections to natural language \cite{KressGazitFP08ar} and other intuitive user interfaces \cite{SrinivasKKKF13robio}.
In order for LTL-based control synthesis methods to move outside research labs and be widely adopted by the robotics community as a specification language of choice, specification debugging tools must be developed as well.
In \cite{KimFS12icra,KimF12iros}, we studied the theoretical foundations of the specification automata revision problem and we proposed heuristic algorithms for its solution.
In \cite{KimF13icra}, we presented a version of the revision problem for weighted transition systems.
In the last formulation, the debugging and revision problem becomes harder to solve since the specification could fail due to not satisfying certain cost constraints, such as, the battery capacity, certain time limit, etc.

Here, we revisit the problem posed in \cite{KimFS12icra}.
When automatically revising specifications, we are often faced with the challenge that not all goals have the same value for the user.
In particular, we assume that the user has certain utility or preference value for each of the subgoals.
Thus, an automatic specification revision should recommend removing the least desirable goals.
In detail, we assume that the specification is provided as an $\omega$-automaton, i.e., a finite automaton with B{\"u}chi acceptance conditions, and that each symbol labeling the transitions has a quantitative preference value (i.e., a positive number).

We formulate two different revision problems.
The first problem concerns removing a set of symbols such that the synthesis problem has now a solution and the sum of the preference levels of the set of removed symbols is minimized.
The second problem again seeks to remove a set of symbols such that the synthesis problem has now a solution; but now the largest preference level of the symbols in the removal set must be minimized.

Not surprisingly the former problem is intractable.
However, interestingly, the latter problem can be solved in polynomial time.
We show how the algorithm that we presented in \cite{KimF12iros} can be modified to provide an exact or approximate solution (depending on the cost function) to the revision problem with preferences in polynomial time.
A practical implication of the results in this paper is that the user can now get an exact solution if the goal is to satisfy as many high preference goals as possible.

{\bf Contributions:}
We define two new versions of the problem of revision under quantitative preferences.
We show that one version can be solved optimally in polynomial time while the other version of the problem is in general intractable.
We provide an exact and an approximate, respectively, polynomial time algorithm based on Dijkstra's algorithm.
Finally, we present some examples and we demonstrate the computational savings of our approximate algorithm over the Brute-Force Search Algorithm that solves the intractable version of the problem exactly.

{\bf Related Research:}
The problem of revising or resolving conflicting LTL specifications has received considerable attention recently.
The closest work to ours is presented in \cite{TumovaEtAl13acc}.
The authors consider a number of high-level requirements in LTL which not all can be satisfied on the system.
Each formula that is satisfied gains some reward.
The goal of their algorithm is to maximize the rewards and, thus, maximize the number of requirements that can be satisfied on the system.
Our problem definition is similar in spirit, but the problem goals are substantially different and the two approaches can be viewed as complementary.
In \cite{TumovaEtAl13acc}, if a whole sub-specification cannot be realized, then it is aborted.
In our case, we try to minimally revise the sub-specification so that it can be partially satisfied.
Another substantial difference is that our proposed solutions can be incorporated directly within the control synthesis algorithm.
Namely, as the algorithm searches for a satisfiable plan, it also creates the graph where the search for the revision will take place.
In \cite{TumovaEtAl13acc}, the graph to be used for the revision must be constructed as a separate step.

The problem of LTL planning with qualitative preferences has been studied in \cite{SonPB12waph,BienvenuFM06kr} (see also the references therein for more research in this direction).
As opposed to revision problem, planning with preferences is based on the fact that there are many satisfiable plans and, thus, the most preferable one should be selected.
For LTL games, LTLMop \cite{RamanK11cav} was developed to debug unrealizable LTL specifications in reactive planning for robotic applications.
The problem of revising LTL specifications on-the-fly as the robot explores its environment is studied in \cite{GuoJD13icra}.

In the context of general planners, the problem of finding good excuses on why the planning failed has been studied in \cite{GobelbeckerEtAl10icaps}.
Over-Subscription Planning (OSP) \cite{Smith04icaps} and Partial Satisfaction Planning (PSP) \cite{BrielEtAl04aaai} are also very related problems.
The aforementioned approaches do not consider extended goals in LTL.



\section{Preliminaries}
\label{sec:problem}

In this paper, we work with discrete abstractions (Finite State Machines) of the continuous robotic control system \cite{FainekosGKGP09automatica}.
Each state of the Finite State Machine (FSM) $\FTS$ is labeled by a number of symbols from a set $\Pi= \{\pi_0,$ $\pi_1,$ $\dots,$ $\pi_n\}$ that represent regions in the configuration space of the robot or, more generally, actions that can be performed by the robot.

\begin{defn} [FSM] A Finite State Machine is a tuple $\FTS = (Q, Q_0, \rightarrow_\FTS, h_\FTS, \Pi)$ where:
$Q$ is a set of states; 
$Q_0 \subseteq Q$ is the set of possible initial states; 
$\rightarrow_\FTS = E \subseteq Q \times Q$ is the transition relation; and, 
$h_\FTS : Q \rightarrow \Pc(\Pi)$ maps each state $q$ to the set of atomic propositions that are true on $q$.
\end{defn}

We define a {\it path} $p : \Ne \rightarrow Q$ on the FSM to be a sequence of states and a {\it trace} to be the corresponding sequence of sets of propositions.
Formally, a path is a function $p : \Ne \rightarrow Q$ such that for each $i\in \Ne$ we have $p(i) \rightarrow_\FTS p(i+1)$ and the trace is the function composition $\bar p = h_\FTS \circ p : \Ne \rightarrow \Pc(\Pi)$.
The language $\Lc(\FTS)$ of $\FTS$ consists of all possible traces.

\begin{ass}
All the states on $\FTS$ are reachable.
\label{ass:reach}
\end{ass}

In this work, we are interested in the specification automata that impose certain requirements on the traces of $\FTS$.
In the following, $\Pc(\Pi)$ denotes the powerset of a set $\Pi$.

\begin{defn}
A specification automaton is a tuple $\ASPEC = (S_{\ASPEC}, s_{0}^{\ASPEC}, \Pc(\Pi), \delta_{\ASPEC}, F_{\ASPEC}, \theta)$ where:
\ifthenelse {\boolean{TECHREP}}
{
\begin{itemize}
\item $S_{\ASPEC}$ is a finite set of states;
\item $s_0^{\ASPEC}$ is the initial state; 
\item $\Pc(\Pi)$ is the input alphabet;
\item $\delta_{\ASPEC} : S_{\ASPEC} \times \Pc(\Pi) \rightarrow \Pc(S_{\ASPEC})$ is a transition function;
\item $F_{\ASPEC} \subseteq S_{\ASPEC}$ is a set of final states; and
\item $\theta : \Pi \times S^2_{\ASPEC} \rightarrow \mathbb{R}_{\geq 0}$ is a preference function.
\end{itemize}
}
{
$S_{\ASPEC}$ is a finite set of states;
$s_0^{\ASPEC}$ is the initial state; 
$\Pc(\Pi)$ is the input alphabet;
$\delta_{\ASPEC} : S_{\ASPEC} \times \Pc(\Pi) \rightarrow \Pc(S_{\ASPEC})$ is a transition function;
$F_{\ASPEC} \subseteq S_{\ASPEC}$ is a set of final states; and
$\theta : \Pi \times S^2_{\ASPEC} \rightarrow \mathbb{R}_{\geq 0}$ is a preference function.
}
\end{defn}

When $s' \in \delta_{\ASPEC}(s,l)$, we also write $s \stackrel{l}{\rightarrow}_{\ASPEC} s'$ or $(s, l, s') \in \rightarrow_{\ASPEC}$.
A {\it run} $r$ of ${\ASPEC}$ is a sequence of states $r : \Ne \rightarrow S_{\ASPEC}$ that occurs under an input trace $\bar p$ taking values in $\Pc(\Pi)$.
That is, for $i = 0$ we have $r(0) = s_{0}^{\ASPEC}$ and for all $i \geq 0$ we have $r(i) \stackrel{\bar p(i)}{\rightarrow}_{\ASPEC} r(i+1)$.  
Let $\lim(\cdot)$ be the function that returns the set of states that are encountered infinitely often in the run $r$ of $\ASPEC$. 
Then, a run $r$ of an automaton $\ASPEC$ over an infinite trace $\bar p$ is {\it accepting} if and only if $\lim(r) \cap F_{\ASPEC} \neq \emptyset$.
This is called a B{\"u}chi acceptance condition.
Finally, we define the language $\Lc(\ASPEC)$ of $\ASPEC$ to be the set of all traces $\bar p$ that have a run that is accepted by $\ASPEC$.

In order to simplify the discussion in Section \ref{sec:specrev}, we will make the following assumption without loss of generality.

\begin{ass}
Between any two states of the specification automaton there exists at most one transition.
\end{ass}

We will also be using the following notations.

\begin{itemize}
\item we define the set $E_{\ASPEC} \subseteq S_{\ASPEC}^2$, such that $(s,s') \in E_{\ASPEC}$ iff $\exists l \in \Pc(\Pi)$ , $s \stackrel{l}{\rightarrow}_{\ASPEC} s'$; and, 
\item we define the function $\lambda_{\ASPEC} : S_{\ASPEC}^2 \rightarrow \Pc(\Pi)$ which maps a pair of states to the label of the corresponding transition, i.e., if $s \stackrel{l}{\rightarrow}_{\ASPEC} s'$, then  $\lambda_{\ASPEC}(s,s') = l$.
\end{itemize}
 
In brief, our goal is to generate paths on $\FTS$ that satisfy the specification $\ASPEC$ \cite{FainekosGKGP09automatica}.
This can be achived by finding accepting runs on the product automaton $\Ac = \FTS \times \ASPEC$.

\begin{defn}
The product automaton $\Ac = \FTS \times \ASPEC$ is the automaton $\Ac = (S_\Ac, s_{0}^{\Ac}, \Pc(\Pi), \delta_\Ac, F_\Ac)$ where:
\begin{itemize}
\item $\Sc_\Ac = Q \times S_{\ASPEC}$,
\item $s_{0}^{\Ac} = \{(q_0,s_{0}^{\ASPEC}) \; | \; q_0 \in Q_0\}$,
\item $\delta_\Ac : S_\Ac \times \Pc(\Pi) \rightarrow \Pc(S_\Ac)$ s.t. $(q_j,s_j)$ $\in$ $\delta_\Ac((q_i,s_i),l)$ iff  $q_i \rightarrow_\FTS q_j$ and $s_j \in \delta_{\ASPEC}(s_i,l)$ with $l \subseteq h_\FTS(q_j)$,
\item $F_\Ac = Q \times F$ is the set of accepting states.
\end{itemize}
\end{defn}

We say that $\ASPEC$ is {\it satisfiable} on $\FTS$ if $\Lc(\Ac) \neq \emptyset$.
Moreover, finding a satisfying path on $\FTS \times \ASPEC$ is an easy algorithmic problem \cite{FainekosGKGP09automatica}.
Each accepting (infinite) run consists of two parts: {\bf prefix:} a part that is executed only once (from an initial state to a final state) and, {\bf lasso:} a part that is repeated infinitely (from a final state back to itself). 
Note that if the prefix or the lasso do not contain a final state,
then the language $\Lc(\Ac)$ is empty.
Namely, the synthesis phase has failed and we cannot find a system behavior that satisfies the specification.

When a specification $\Bc$ is not satisfiable on a particular system $\FTS$, the current motion planning  and control synthesis methods based on automata theoretic concepts \cite{FainekosGKGP09automatica,UlusoyEtAl2011iros,LaViersEtAl2011iccps} simply return that the specification is not satisfiable without any other user feedback. 
In such cases, our previous algorithms \cite{KimFS12icra,KimF12iros} can provide as feedback to the user the closest revision under equal preference for all goals.
Formally, a revision $R$ is a subset of $\Pc(\Pi)\times E_{\ASPEC}$.
Each $(\pi,s,s') \in R$ indicates that $\pi$ must be removed from $\lambda_{\ASPEC}(s,s')$.

\section{Revision Under Preferences}
\label{sec:specrev}

When choosing an alternative plan, each user can have different preferences.
Suppose that users can assign some preference level to each proposition labeling the specification automaton through the preference function $\theta$.
When preference level is 0, it is least preferred, and the greater preference level is, the more preferred it is. However, preference level cannot be $\infty$.
We remark that each occurrence of an atomic proposition over different transitions can have different preference levels.
Therefore, taking transitions on the cross-product automaton $\Ac$, we can get as a reward preference levels of elements in $\Pi$ on the transitions.


A revised specification is one that can be satisfied on the discrete abstraction of the workspace or the configuration space of the robot.
In order to search for a minimal revision, we need first to define an ordering relation on automata as well as a distance function between automata.
We do not want to consider the ``space" of all possible automata, but rather the ``space" of specification automata which are semantically close to the initial specification automaton $\ASPEC$.
The later will imply that we remain close to the initial intention of the designer.
We propose that this space consists of all the automata that can be derived from $\ASPEC$ by removing symbols from the transitions.
Our definition of the ordering relation between automata relies upon the previous assumption.

\begin{defn}[Relaxation]
Let $\BUCHI_1 = (S_{\BUCHI_1}$, $s_0^{\BUCHI_1}$, $\Pc(\Pi)$, $\rightarrow_{\BUCHI_1}$, $F_{\BUCHI_1}$, $\theta_{\BUCHI_1})$ and $\BUCHI_2 = (S_{\BUCHI_2}, s_{0}^{\BUCHI_2}, \Pc(\Pi), \rightarrow_{\BUCHI_2}, F_{\BUCHI_2}, \theta_{\BUCHI_2})$  be two specification automata having the same preference levels for $\Pc(\Pi)$.
Then, we say that $\BUCHI_2$ is a relaxation of $\BUCHI_1$ and we write $\BUCHI_1 \preceq \BUCHI_2$ if and only if $S_{\BUCHI_1} = S_{\BUCHI_2} = S$, $s_{0}^{\BUCHI_1} = s_{0}^{\BUCHI_2}$, $F_{\BUCHI_1} = F_{\BUCHI_2}$, $\theta_{\BUCHI_1} = \theta_{\BUCHI_2}$ and 
\begin{enumerate}
\item $\forall (s,l,s') \in \rightarrow _{\BUCHI_1} - \rightarrow _{\BUCHI_2}$ . $\exists l'$ . 

$(s,l',s') \in \rightarrow_{\BUCHI_2} - \rightarrow_{\BUCHI_1}$ and $l' \subseteq l$. 
\item $\forall (s,l,s') \in \rightarrow _{\BUCHI_2} - \rightarrow _{\BUCHI_1}$ . $\exists l'$ . 

$(s,l',s') \in \rightarrow_{\BUCHI_1} - \rightarrow_{\BUCHI_2}$ and $l \subseteq l'$. 
\end{enumerate}
\label{def:valid_rel_B}
\end{defn}

We remark that if $\BUCHI_1 \preceq \BUCHI_2$, then $\Lc(\BUCHI_1) \subseteq \Lc(\BUCHI_2)$ since the relaxed automaton allows more behaviors to occur.

We can now define the set of automata over which we will search for a revision.

\begin{defn}
Given a system $\FTS$ and and a specification automaton $\ASPEC$, the set of {\it valid relaxations} of $\ASPEC$ is defined as 
$\Rr(\ASPEC,\FTS) = \{ \BUCHI \; | \; \ASPEC \preceq \BUCHI \mbox{ and } \Lc(\FTS \times \BUCHI) \neq \emptyset \}.$
\end{defn}

We can now search for a solution in the set $\Rr(\ASPEC,\FTS)$.
Different solutions can be compared from their revision sets.

\begin{defn}[Revision Set]
Given a specification automaton $\ASPEC$ and a $\BUCHI \in \Rr(\ASPEC,\FTS)$, the revision set is defined as $R(\ASPEC,\BUCHI) = \{ (\pi,s,s') \; | \; \pi \in (\lambda_{\ASPEC}(s,s') - \lambda_{\BUCHI}(s,s')) \}$.
\label{defn:rev:set}
\end{defn}

We define two different revision problems.

\begin{prob}[Min-Sum Revision]
Given a system $\FTS$ and a specification automaton $\ASPEC$, if the specification $\ASPEC$ is not satisfiable on $\FTS$, then find a revision set $R$ such that $\sum_{\rho\in R} \theta(\rho)$ is minimized.
\label{prb:planning:sum}
\end{prob}

\begin{prob}[Min-Max Revision]
Given a system $\FTS$ and a specification automaton $\ASPEC$, if the specification $\ASPEC$ is not satisfiable on $\FTS$, then find a revision set $R$ such that $\max_{\rho\in R} \theta(\rho)$ is minimized.
\label{prb:planning:max}
\end{prob}


The edges of $G_\Ac$ are labeled by the set of symbols which if removed from the corresponding transition on $\ASPEC$, they will enable the transition on $\Ac$.
The overall problem then becomes one of finding the least number of symbols to be removed in order for the product graph to have an accepting run. 

\begin{defn}
  Given a system $\FTS$ and a specification automaton $\ASPEC$, we
  define the graph $G_{\Ac} = (V,E,v_s,V_f,\APRem,\Lambda,p)$, which corresponds to
  the product $\Ac = \FTS \times \ASPEC$ as follows
\begin{itemize}
\item $V = \Sc$ is the set of nodes
\item $E = E_\Ac \cup E_D \subseteq \Sc \times \Sc$, where $E_\Ac$ is
  the set of edges that correspond to transitions on $\Ac$, i.e.,
  $((q,s),(q',s')) \in E_\Ac$ iff $\exists l \in \Pc(\Pi)$ . $(q,s)
  \stackrel{l}{\rightarrow}_\Ac (q',s')$; and $E_D$ is the set of
  edges that correspond to disabled transitions, i.e.,
  $((q,s),(q',s')) \in E_D$ iff $q \rightarrow_\FTS q'$ and
  $s \stackrel{l}{\rightarrow}_{\ASPEC} s'$ with $l \cap
  (\Pi-h_\FTS(q'))\neq \emptyset$
\item $v_s = s_0^{\Ac}$ is the source node
\item $V_f = F_\Ac$ is the set of sinks
\item $\APRem = \{ \tupleof{\pi,(s,s')} \; | \; \pi \in \Pi, (s,s') \in E_{\ASPEC} \}$
\item $\Lambda : E \rightarrow \Pc(\APRem)$ is the edge labeling function such that if $e = ((q,s),
(q',s'))$, then 
\[\Lambda(e) = \{ \tupleof{\pi,(s,s')} \; | \; \pi \in (\lambda_{\ASPEC}(s,s') - h_\FTS(q')) \}.\]
\item $\theta : \APRem \rightarrow \mathbb{R}_{\ge 0}$ is the preference function of $\ASPEC$ restricted on $\APRem$.
\end{itemize}
\label{def:graph}
\end{defn}

If $\Lambda(e) \not= \emptyset$, then it specifies those atomic propositions in $\lambda_{\ASPEC}(s,s')$ that need to be removed in order to enable the edge in $\Ac$.
Again, note that the labels of the edges of $G_\Ac$ are subsets of $\APRem$ rather than $\Pi$.
This is due to the fact that we are looking into removing an atomic proposition $\pi$ from a specific transition $(s,l,s')$ of $\ASPEC$ rather than all occurrences of $\pi$ in $\ASPEC$.

Consider now a path that reaches an accept state and then can loop back to the same accept state.
The set of labels of the path is a revision set $R$ that corresponds to some $\BUCHI \in  \Rr(\ASPEC,\FTS)$.
This is immediate by the definition of the graph $G_\Ac$.
Thus, our goal is to solve the Min-Sum and Min-Max revision problems on this graph.

\ifthenelse {\boolean{TECHREP}}
{
\ifthenelse {\boolean{BGRAPHPDF}}
{
\begin{figure}[t]
\centering
\includegraphics[width=8cm]{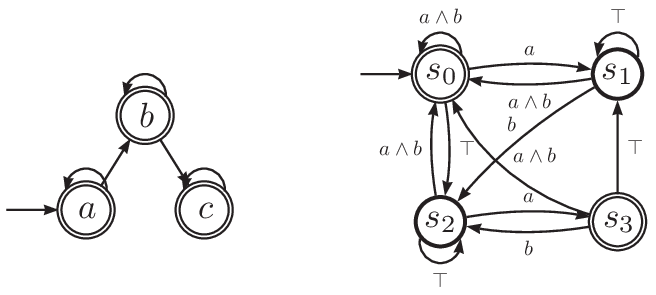}
\caption{The system $\FTS$ and the specification $\ASPEC$ of Example \ref{exmp:spec:01}. The LTL formula of $\ASPEC$ is $GF(a \land Fb)$.}
\label{fig:exmp:spec:01:system}
\end{figure}
}
{
\begin{figure}
\begin{center}
\VCDraw{%
\begin{VCPicture}{(0,-1.7)(9,2)}
\FinalState[a]{(0,-0.8)}{A} \FinalState[b]{(1,0.8)}{B} \FinalState[c]{(2,-0.8)}{C}
\Initial{A}
\ncline{->}{A}{B} \naput{}
\ncline{->}{B}{C} \naput{}
\nccircle{->}{A}{0.35cm} \naput{}
\nccircle{->}{B}{0.35cm} \naput{}
\nccircle{->}{C}{0.35cm} \naput{}

\FinalState[s_0]{(6,1.5)}{0}
\State[s_1]{(9,1.5)}{1}
\State[s_2]{(6,-1.0)}{2}
\FinalState[s_3]{(9,-1.0)}{3}
\Initial{0}
\ncarc[arcangle=10]{->}{0}{1} \naput{$a$}
\ncarc[arcangle=10]{->}{1}{0} \naput{$a \land b$}
\ncarc[arcangle=10]{->}{0}{2} \naput{$\top$}
\ncarc[arcangle=10]{->}{2}{0} \naput{$a \land b$}
\ncarc[arcangle=10]{->}{2}{3} \naput{$a$}
\ncarc[arcangle=10]{->}{3}{2} \naput{$b$}
\ncarc[arcangle=25]{->}{3}{0} \nbput{$a \land b$}
\ncarc[arcangle=-10]{->}{1}{2} \nbput{$b$}
\ncline{->}{3}{1} \nbput{$\top$}
\nccircle{->}{0}{0.35cm}\nbput{$a \land b$}
\nccircle{->}{1}{0.35cm}\nbput{$\top$}
\nccircle{->}{2}{-0.35cm}\naput{$\top$}
\end{VCPicture}}
\end{center}
\caption{The system $\FTS$ and the specification $\ASPEC$ of Example \ref{exmp:spec:01}. The LTL formula of $\ASPEC$ is $GF(a \land Fb)$.}
\label{fig:exmp:spec:01:system}
\end{figure}
}

\ifthenelse {\boolean{BGRAPHPDF}}
{
\begin{figure}[t]
\centering
\includegraphics[width=8cm]{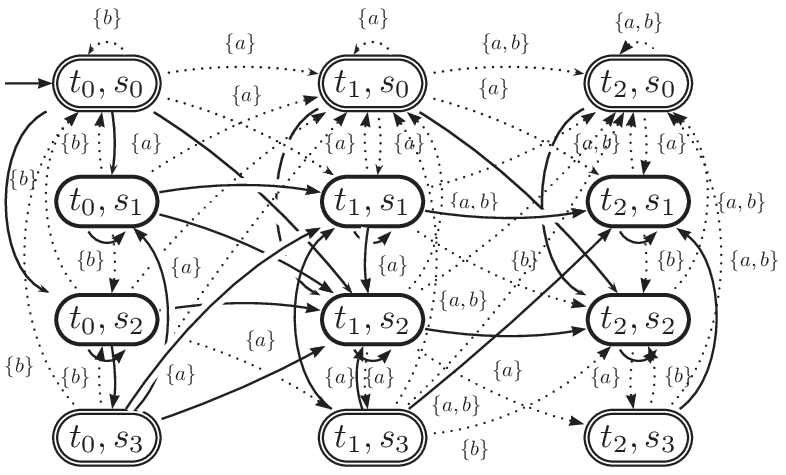}
\caption{The cross-product automaton $\FTS \times \ASPEC$ with relaxations. Solid transition are for valid transitions and dotted transitions are for relaxed transitions.}
\label{fig:exmp:spec:01:cross}
\end{figure}
}
{
\begin{figure}
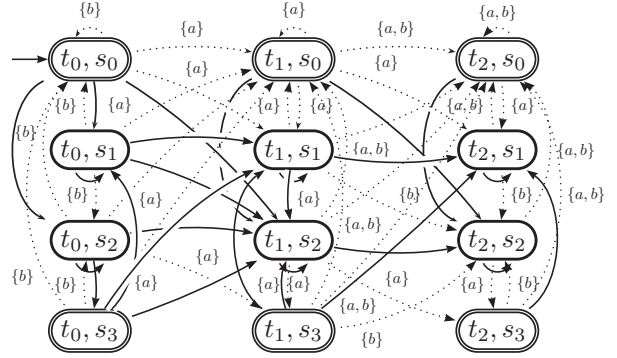

\begin{center}
\VCDraw{%
\begin{VCPicture}{(0,-3.2)(9,3.8)}
\StateLineDouble \StateVar[t_0,s_0]{(0,3)}{t0s0}
\StateLineDouble \StateVar[t_1,s_0]{(4.5,3)}{t1s0}
\StateLineDouble \StateVar[t_2,s_0]{(9,3)}{t2s0}
\StateLineSimple \StateVar[t_0,s_1]{(0,1)}{t0s1}
\StateLineSimple \StateVar[t_1,s_1]{(4.5,1)}{t1s1}
\StateLineSimple \StateVar[t_2,s_1]{(9,1)}{t2s1}
\StateLineSimple \StateVar[t_0,s_2]{(0,-1)}{t0s2}
\StateLineSimple \StateVar[t_1,s_2]{(4.5,-1)}{t1s2}
\StateLineSimple \StateVar[t_2,s_2]{(9,-1)}{t2s2}
\StateLineDouble \StateVar[t_0,s_3]{(0,-3)}{t0s3}
\StateLineDouble \StateVar[t_1,s_3]{(4.5,-3)}{t1s3}
\StateLineDouble \StateVar[t_2,s_3]{(9,-3)}{t2s3}

\nccircle[doubleline=false,linestyle=dotted]{->}{t0s0}{0.35cm} \nbput{$\{b\}$}
\ncarc[doubleline=false,linestyle=dotted,arcangle=10]{->}{t0s0}{t1s0} \naput{$\{a\}$}
\ncarc[doubleline=false,arcangle=10]{->}{t0s0}{t0s1}
\ncarc[doubleline=false,arcangle=-65]{->}{t1s0}{t1s2}
\ncarc[doubleline=false,linestyle=dotted,arcangle=10,border=0.1cm]{->}{t0s0}{t1s1} \naput[npos=.4]{$\{a\}$}
\ncarc[doubleline=false,arcangle=-65,border=0.1cm,nodesep=0.1cm]{->}{t0s0}{t0s2}
\ncarc[doubleline=false,arcangle=10,border=0.1cm,nodesep=0]{->}{t0s0}{t1s2}

\nccircle[doubleline=false,linestyle=dotted]{->}{t1s0}{0.35cm} \nbput{$\{a\}$}
\ncarc[doubleline=false,linestyle=dotted,arcangle=10]{->}{t1s0}{t2s0} \naput{$\{a,b\}$}
\ncarc[doubleline=false,linestyle=dotted,arcangle=10]{->}{t1s0}{t1s1} \naput{$\{a\}$}
\ncarc[doubleline=false,arcangle=-65]{->}{t2s0}{t2s2}
\ncarc[doubleline=false,linestyle=dotted,arcangle=10]{->}{t1s0}{t2s1} \naput[npos=.3]{$\{a\}$}
\ncarc[doubleline=false,arcangle=10]{->}{t1s0}{t2s2}
\Initial{t0s0}

\nccircle[doubleline=false,linestyle=dotted]{->}{t2s0}{0.35cm} \nbput{$\{a,b\}$}
\ncarc[doubleline=false,linestyle=dotted,arcangle=10]{->}{t2s0}{t2s1} \naput{$\{a\}$}
\ncarc[doubleline=false,linestyle=dotted,arcangle=10]{->}{t0s1}{t0s0} \naput{$\{b\}$}
\ncarc[doubleline=false,linestyle=dotted,arcangle=10]{->}{t0s1}{t1s0} \naput[npos=.1]{$\{a\}$}

\nccircle[doubleline=false]{->}{t0s1}{-0.35cm}
\ncarc[doubleline=false,arcangle=10,border=0.1cm]{->}{t0s1}{t1s1}
\ncarc[doubleline=false,linestyle=dotted,arcangle=10]{->}{t0s1}{t0s2} \nbput[npos=.5]{$\{b\}$}
\ncarc[doubleline=false,arcangle=10,border=0.1cm]{->}{t0s1}{t1s2}
\ncarc[doubleline=false,linestyle=dotted,arcangle=10]{->}{t1s1}{t1s0} \naput[npos=.5]{$\{a\}$}
\ncarc[doubleline=false,linestyle=dotted,arcangle=-10]{->}{t1s1}{t2s0} \nbput[npos=.2]{$\{a,b\}$}

\nccircle[doubleline=false]{->}{t1s1}{-0.35cm}
\ncarc[doubleline=false,arcangle=-10,border=0.1cm]{->}{t1s1}{t2s1}
\ncarc[doubleline=false,arcangle=-10,border=0.1cm]{->}{t1s1}{t1s2}
\ncarc[doubleline=false,linestyle=dotted,arcangle=-10]{->}{t1s1}{t2s2} \naput[npos=.6]{$\{b\}$}
\ncarc[doubleline=false,linestyle=dotted,arcangle=10]{->}{t2s1}{t2s2} \naput[npos=.5]{$\{b\}$}
\ncarc[doubleline=false,linestyle=dotted,arcangle=10]{->}{t2s1}{t2s0} \naput[npos=.5]{$\{a,b\}$}

\nccircle[doubleline=false]{->}{t2s1}{-0.35cm}
\ncarc[doubleline=false,linestyle=dotted,arcangle=45]{->}{t0s2}{t0s0} \naput[npos=.6]{$\{b\}$}
\ncarc[doubleline=false,linestyle=dotted,arcangle=10]{->}{t0s2}{t1s0} \nbput[npos=.2]{$\{a\}$}

\nccircle[doubleline=false]{->}{t0s2}{-0.35cm}
\ncarc[doubleline=false,arcangle=10]{->}{t0s2}{t1s2}
\ncarc[doubleline=false,arcangle=10]{->}{t0s2}{t0s3}
\ncarc[doubleline=false,linestyle=dotted,arcangle=10]{->}{t0s2}{t1s3} \naput[npos=.5]{$\{a\}$}
\ncarc[doubleline=false,linestyle=dotted,arcangle=-50]{->}{t1s2}{t1s0} \naput[npos=.1]{$\{a\}$}
\ncarc[doubleline=false,linestyle=dotted,arcangle=-10]{->}{t1s2}{t2s0} \nbput[npos=.1]{$\{a,b\}$}

\nccircle[doubleline=false]{->}{t1s2}{-0.35cm}
\ncarc[doubleline=false,arcangle=-10]{->}{t1s2}{t2s2}
\ncarc[doubleline=false,linestyle=dotted,arcangle=-10]{->}{t1s2}{t1s3} \nbput[npos=.5]{$\{a\}$}
\ncarc[doubleline=false,linestyle=dotted,arcangle=-10]{->}{t1s2}{t2s3} \naput[npos=.5]{$\{a\}$}

\nccircle[doubleline=false]{->}{t2s2}{-0.35cm}
\ncarc[doubleline=false,linestyle=dotted,arcangle=-50]{->}{t2s2}{t2s0} \nbput[npos=.5]{$\{a,b\}$}
\ncarc[doubleline=false,linestyle=dotted,arcangle=-10]{->}{t2s2}{t2s3} \nbput[npos=.5]{$\{a\}$}
\ncarc[doubleline=false,linestyle=dotted,arcangle=45]{->}{t0s3}{t0s0} \naput[npos=.2]{$\{b\}$}
\ncarc[doubleline=false,linestyle=dotted,arcangle=5]{->}{t0s3}{t1s0} \nbput[npos=.15]{$\{a\}$}
\ncarc[doubleline=false,arcangle=-45,border=0.1cm]{->}{t0s3}{t0s1}
\ncarc[doubleline=false,arcangle=15,border=0.1cm]{->}{t0s3}{t1s1}
\ncarc[doubleline=false,linestyle=dotted,arcangle=10]{->}{t0s3}{t0s2} \naput[npos=.5]{$\{b\}$}
\ncarc[doubleline=false,arcangle=-5,border=0.1cm]{->}{t0s3}{t1s2}

\ncarc[doubleline=false,linestyle=dotted,arcangle=-35]{->}{t1s3}{t1s0} \naput[npos=.08]{$\{a\}$}
\ncarc[doubleline=false,linestyle=dotted,arcangle=-10]{->}{t1s3}{t2s0} \nbput[npos=.08]{$\{a,b\}$}
\ncarc[doubleline=false,arcangle=55]{->}{t1s3}{t1s1}
\ncarc[doubleline=false,arcangle=-3]{->}{t1s3}{t2s1}
\ncarc[doubleline=false,arcangle=20]{->}{t1s3}{t1s2}
\ncarc[doubleline=false,linestyle=dotted,arcangle=-20]{->}{t1s3}{t2s2} \nbput[npos=.2]{$\{b\}$}

\ncarc[doubleline=false,linestyle=dotted,arcangle=-20]{->}{t2s3}{t2s2} \nbput[npos=.5]{$\{b\}$}
\ncarc[doubleline=false,linestyle=dotted,arcangle=-45]{->}{t2s3}{t2s0} \nbput[npos=.5]{$\{a,b\}$}
\ncarc[doubleline=false,arcangle=-55]{->}{t2s3}{t2s1}
\end{VCPicture}}
\end{center}
\caption{The cross-product automaton $\FTS \times \ASPEC$ with relaxations. Solid transition are for valid transitions and dotted transitions are for relaxed transitions.}
\label{fig:exmp:spec:01:cross}
\end{figure}
}

\begin{exmp}
Let us consider the system $\FTS$ in Fig \ref{fig:exmp:spec:01:system}.
The LTL formula of the specification $\ASPEC$ in Fig. \ref{fig:exmp:spec:01:system} is $GF(a \land Fb)$\footnote{For LTL semantics please see \cite{FainekosGKGP09automatica}. In this paper, we use LTL formulas only as notational convenience to represent larger automata.}.
Informally, the specification is `Infinitely often visit a and then visit b'. Fig. \ref{fig:exmp:spec:01:cross} is the cross-product automaton $\FTS \times \ASPEC$. The initial state of the cross-product automaton is $(t_0,s_0)$. The final states are $(t_0,s_0), (t_1,s_0), (t_2,s_0), (t_0,s_3), (t_1,s_3), (t_2,s_3)$.
$\ASPEC$ is not satisfiable on $\FTS$ so that there is no reachable path from the state $(t_0,s_0)$ to one of the finals and from one of the final states to back to itself.
In this example, the set of atomic propositions is $\Pi = \{a, b, c\}$.
Suppose that the preference levels of the atomic propositions are $\theta((s_i,s_j),\{a\})=3$, $\theta((s_i,s_j),\{b\})=5$, $\theta((s_i,s_j),\{c\})=4$ where $\forall s_i, s_j \in S_{\BUCHI}$.
Then from valid relaxations of $\ASPEC$, we can find acceptable paths as follows:
$p_1 = \langle$$((t_0,s_0)$, $\{b\}$, $(t_0,s_0))$ $((t_0,s_0)$, $\{b\}$, $(t_0,s_0))$ $\ldots$$\rangle$, 
$p_2 = \langle$$((t_0,s_0)$, $\emptyset$, $(t_0,s_1))$ $((t_0,s_1)$, $\{b\}$, $(t_0,s_0))$ $((t_0,s_0)$, $\emptyset$, $(t_0,s_1))$ $\ldots$$\rangle$, 
$p_3 = \langle$$((t_0,s_0)$, $\{a\}$, $(t_1,s_0))$ $((t_1,s_0)$, $\{a\}$, $(t_1,s_0))$ $((t_1,s_0)$, $\{a\}$, $(t_1,s_0)) \ldots$$\rangle$,
$p_4 = \langle$$((t_0,s_0)$, $\{a\}$, $(t_1,s_0))$ $((t_1,s_0)$, $\{a\}$, $(t_1,s_0))$ $((t_1,s_0)$, $\{a,b\}$, $(t_2,s_0))$ $((t_2,s_0)$, $\{a,b\}$, $(t_2,s_0)) \ldots$$\rangle$, etc.
The sum of preference levels of each path are 5, 5, 3, 8, respectively. The max of preference levels of each path are 5, 5, 3, 5.
Therefore, among the above paths, the path having atomic propositions that minimize the sum of preference levels is $p_3$. It has only $\{a\}$ on the transitions, so the sum of preference level of the path is 3 and the max of preference level of the path is also 3.
\label{exmp:spec:01}
\exmend
\end{exmp}

}
{
}

First, we study the computational complexity of the two problems by restricting the search problem only to paths from source (initial state) to sink (accept state).
Let $Paths(G_\Ac)$ denote all such paths on $G_\Ac$.
We indicate that the graph search equivalent problem of Problem \ref{prb:planning:max} is in P.
Given a path $p = v_s v_1 v_2 \ldots v_f$ on $G_\Ac$ with $v_f \in V_f$, we define the max-preference level of the path to be:
\[
\theta_{\max}(p) = \max_{(v_i,v_{i+1})\in p} \theta(\Lambda(v_i,v_{i+1}))
\]
Note that this is the same as the original cost function in Problem \ref{prb:planning:max} since clearly $\max_{(v_i,v_{i+1})\in p} \theta(\Lambda(v_i,v_{i+1})) = \max_{\rho \in R} \theta(\rho)$ where $R = \cup_{(v_i,v_{i+1})\in p}\Lambda(v_i,v_{i+1})$.
Thus, Problem \ref{prb:planning:max} is converted into the following optimization problem:
\begin{equation}
\label{eq:min:max}
p^* = \arg\min_{p \in \mbox{$Paths$}(G_\Ac)} \theta(p)
\end{equation}
And, thus, the revision will be $R = \cup_{(v_i,v_{i+1})\in p^*}\Lambda(v_i,v_{i+1})$.
Now, we recall the weak optimality principle \cite{MartinsPRDS99}.

\begin{defn}[Weak optimality principle]
There is an optimal path formed by optimal subpaths.
\end{defn}

\begin{prop}
The graph search equivalent of Problem \ref{prb:planning:max} satisfies the weak optimality principle.
\end{prop}

\ifthenelse {\boolean{TECHREP}}
{
\begin{proof}
Let $p^*$ be an optimal path under the cost function $\theta_{\max}$, that is, for any other path $p$, we have $\theta_{\max}(p)\geq \theta_{\max}(p*)$.
We assume that $p^*$ is a loopless path.
Notice if a loop exists, then it can be removed without affecting the cost of the path.
Let $p^*$ have a subpath $p_s = v_1 v_2 \ldots v_{i-1} v_i$ which is not optimal, that is $p* = p_1 \circ p_s \circ p_2$.
We use here the notation $p_1\circ p_2$ to indicate that the last vertex of $p_1$ and the first vertex of $p_2$ are the same and are going to be merged.
Now assume that there is another subpath $p'_s = v_1 v'_2 \ldots v'_{j-1} v_i$ such that $\theta_{\max}(p_s) > \theta_{\max}(p'_s)$.
Note that $\theta_{\max}(p_s) \leq \theta_{\max}(p_1)$ and $\theta_{\max}(p_s) \leq \theta_{\max}(p_2)$ otherwise $p^*$ would not be optimal.
We have $\theta_{\max}(p_s) = \max(\theta_{\max}(p_1),\theta_{\max}(p_s),\theta_{\max}(p_2)) = \max(\theta_{\max}(p_1),\theta_{\max}(p'_s),\theta_{\max}(p_2)) = \theta_{\max}(p_1\circ p'_s \circ p_2)$.
Hence, the path $p_1\circ p'_s \circ p_2$ is also optimal.
If this process is repeated, we can construct an optimal path $p^{**}$ that contains only optimal subpaths.
\end{proof}
}
{
}

The importance of the weak optimality principle being satisfied is that label correcting and label setting algorithms can be applied to such problems \cite{MartinsPRDS99}.
Dijkstra's algorithm is such an algorithm \cite{LaValle05} and, thus, it can provide an exact solution to the problem.

Now, we proceed to the Min-Sum preference problem.
Given a path $p = v_s v_1 v_2 \ldots v_f$ on $G_\Ac$ with $v_f \in V_f$, we define the sum-preference level of the path to be:
\[
\theta_{+}(p) =  \sum \{ \theta(\rho) \; | \; \rho \in \cup_{(v_i,v_{i+1})\in p}\Lambda(v_i,v_{i+1}) \}
\]
and if we are directly provided with a revision set $R$, then
\[
\theta_{+}(R) =  \sum_{\rho\in R} \theta(\rho) 
\]

\begin{prob}
Labeled Path under Additive Preferences (LPAP). 
\text{\sc Inputs:} A graph $G_{\Ac} = (V,E,v_s,V_f,\APRem,\Lambda,\theta)$,
and a preference bound $K \in \Ne$.
\text{\sc Output:} a set $R \subseteq \APRem$ such that removing all elements in $R$ from edges in $E$ enables a path from $v_s$ to some final vertex $v_f \in V_f$ and $\theta_+(R) \leq K$.
\end{prob}


We can show that the corresponding decision problem is NP-Complete.

\begin{thm}
Given an  instance of the LPAP $(G_{\Ac}, K)$, the decision problem of whether there exists a path $p$ such that $\theta_+(p) \leq K$ is NP-Complete.
\end{thm}

\ifthenelse {\boolean{TECHREP}}
{
\begin{proof}[Sketch]
Clearly, the problem is in NP since given a sequence of nodes $p$, we can verify in polynomial time that $p$ is a path on $G_\Ac$ and $\theta_+(p) \leq K$.

The problem is NP-hard since we can easily reduce the revision problem without preferences (see \cite{KimFS12icra}) to this one by setting the preference levels of all atomic propositions equal to 1. 
Then, since all atomic propositions have the same preference level which is 1, it becomes the problem to find the minimal number of atomic propositions of the graph.
\end{proof}
}
{
}

\ifthenelse {\boolean{TECHREP}}
{

\section{Algorithms for the Revision Problem with Preferences}
\label{sec:heuristic}

In this section, we present Algorithms for the Revision Problem with Preferable (ARPP).
It is based on the Approximation Algorithm of the Minimal Revision Problem (AAMRP) \cite{KimF12iros} which is in turn based on Dijkstra's shortest path algorithm \cite{LaValle05}.
The main difference from AAMRP is that instead of finding the minimum number of atomic propositions that must be removed from each edge on the paths of the graph $G_\Ac$,
ARPP tracks paths having atomic propositions that minimize the preferable level from each edge on the paths of the graph $G_\Ac$.

Here, we present the pseudocode for ARPP.
ARPP is similar to AAMRP in \cite{KimF12iros}.
The difference from \cite{KimF12iros} is that AARP uses \text{\sc Pref} function instead of using cardinality of the set.
For Min-Sum Revision, the function \text{\sc Pref}$:\APRem \rightarrow \mathbb{R}_{\ge 0}$ is defined as following: given a set of label $R \subseteq \APRem$ and the preference function $\theta_{+}: \APRem \rightarrow \mathbb{R}_{\ge 0}$,
\[ \text{\sc Pref}(R) = \theta_{+}(R).\]
The Min-Sum ARPP is denoted by $ARPP_{+}$.

For Min-Max Revision, the function \text{\sc Pref}$:\APRem \rightarrow \mathbb{R}_{\ge 0}$ is defined as following: given a set of label $R \subseteq \APRem$ and the preference function $\theta: \APRem \rightarrow \mathbb{R}_{\ge 0}$,
\[ \text{\sc Pref}(R) = \max_{\rho \in R} \theta(\rho).\]
The Min-Max ARPP is denoted by $ARPP_{max}$.

The main algorithm (Alg. \ref{alg:haprp:main}) divides the problem into two tasks.
First, in line \ref{alg:main:prefix}, it finds an approximation to the minimum preference level of atomic propositions from $\APRem$ that must be removed to have a prefix path to each reachable sink (see Section \ref{sec:problem}).
Then, in line \ref{alg:main:lasso}, it repeats the process from each reachable final state to find an approximation to the minimum preference level of atomic propositions from $\APRem$ that must be removed so that a lasso path is enabled.
The combination of prefix/lasso that removes the least preferable atomic propositions is returned to the user.

\ifthenelse {\boolean{BGRAPHPDF}}
{
\begin{algorithm}[tb]
{\bf Inputs}: a graph $G_\Ac = (V,E,v_s,V_f,\APRem,\Lambda,p)$. \\
{\bf Outputs}: the list $L$ of atomic propositions form $\APRem$ that must be removed $\ASPEC$. 
\caption{ARPP}
\label{alg:haprp:main}
\begin{algorithmic}[1]
\Procedure{ARPP}{$G_\Ac$}
\State $L \gets \APRem$
\State \Comment{Each row of $\Mc$ is set to $(\APRem,\infty)$}
\State $\Mc[:,:] \gets (\APRem,\infty)$
\State $\Mc[v_s,:] \gets (\emptyset,0)$ \Comment{Initialize the source node}
\State $\tupleof{\Mc, \Pb, \Vc} \gets  \text{\sc FindMinPath}(G_\Ac,\Mc,0)$ \label{alg:main:prefix}
\If {$\Vc \cap V_f = \emptyset$}
\State $L \gets \emptyset$
\Else
\For {$v_f \in \Vc \cap V_f$} \label{alg:main:loop}
\State $L_p \gets \text{\sc GetAPFromPath}(v_s,v_f,\Mc,\Pb)$  \label{alg:main:lasso}
\State $\Mc'[:,:] \gets (\APRem,\infty)$
\State $\Mc'[v_f,:] \gets \Mc[v_f,:]$
\State $G_\Ac' \gets (V,E,v_f,\{v_f\},\APRem,L)$
\State $\tupleof{\Mc', \Pb', \Vc'} \gets  \text{\sc FindMinPath}(G'_\Ac,\Mc',1)$
\If {$v_f \in \Vc'$}
\State $L_l \gets \text{\sc GetAPFromPath}(v_f,v_f,\Mc',\Pb')$
\If {\text{\sc Pref}$(L_p \cup L_l) \leq \text{\sc Pref}(L)$}
\State $L \gets L_p \cup L_l$ 
\EndIf
\EndIf
\EndFor
\EndIf
\State \Return $L$
\EndProcedure
\end{algorithmic}
The function \text{\sc GetAPFromPath}($(v_s,v_f,\Mc,\Pb)$) returns the atomic propositions that must be removed from $\ASPEC$ in order to enable a path on $\Ac$ from a starting state $v_s$ to a final state $v_f$ given the tables $\Mc$ and $\Pb$.
\end{algorithm}
}
{
}

Algorithm \ref{alg:haprp:sub} follows closely Dijkstra's shortest path algorithm \cite{CormenLRS01}. 
It maintains a list of visited nodes $\Vc$ and a table $\Mc$ indexed by the graph vertices which stores the set of atomic propositions that must be removed in order to reach a particular node on the graph.
Given a node $v$, the preference level of the set $\Mc[v,1]$ is an upper bound on the minimum preference level of atomic propositions that must be removed.
That is, if we remove all $\overline{\pi} \in \Mc[v,1]$ from $\ASPEC$, then we enable a simple path (i.e., with no cycles) from a starting state to the state $v$. 
The preference level of $|\Mc[v,1]|$ is stored in $\Mc[v,2]$ which also indicates that the node $v$ is reachable when $\Mc[v,2] < \infty$.

The algorithm works by maintaining a queue with the unvisited nodes on the graph.
Each node $v$ in the queue has as key the summed preference level of atomic propositions that must be removed so that $v$ becomes reachable on $\Ac$.
The algorithm proceeds by choosing the node with the minimally summed preference level of atomic propositions discovered so far (line \ref{alg:haprp:sub:min}).
Then, this node is used in order to updated the estimates for the minimum preference level of atomic propositions needed in order to reach its neighbors (line \ref{alg:haprp:sub:relax}).
A notable difference of Alg. \ref{alg:haprp:sub} from Dijkstra's shortest path algorithm is the check for lasso paths in lines \ref{alg:haprp:sub:1}-\ref{alg:haprp:sub:end}.
After the source node is used for updating the estimates of its neighbors, its own estimate for the minimum preference level of atomic propositions is updated either to the value indicated by the self loop or the maximum possible preference level of atomic propositions.
This is required in order to compare the different paths that reach a node from itself.

\ifthenelse {\boolean{BGRAPHPDF}}
{
\begin{algorithm}[tb]
\caption{{\sc FindMinPath}}
{\bf Inputs}: a graph $G_\Ac = (V,E,v_s,V_f,\APRem,\Lambda,p)$, a table
$\Mc$ and a flag $lasso$ on whether this is a lasso path search. \\
{\bf Variables}: a queue $\Qc$, a set $\Vc$ of visited nodes and a
table $\mathbf{P}$ indicating the parent of each node on a path. \\
{\bf Output}: the tables $\Mc$ and $\mathbf{P}$ and the visited nodes $\Vc$
\label{alg:haprp:sub}
\begin{algorithmic}[1]
\Procedure{{\sc FindMinPath}}{$G_\Ac$,$\Mc$,$lasso$}
\State $\Vc \gets \{v_s\}$
\State $\mathbf{P}[:]  \gets \emptyset$ \Comment{Each entry of
$\mathbf{P}$ is set to $\emptyset$}
\State $\Qc \gets V - \{v_s\}$
\For {$v \in V$ such that $(v_s,v) \in E$ and $v \neq v_s$}
\label{alg:haprp:sub:for1}
\State $\tupleof{\Mc,\Pb} \gets \text{\sc Relax}((v_s,v),\Mc,\Pb,\Lambda)$
\EndFor
\If {$lasso=1$} \label{alg:haprp:sub:1}
\If {$(v_s,v_s) \in E$}
\State $\Mc[v_s,1] \gets \Mc[v_s,1] \cup \Lambda(v_s,v_s)$
\State $\Mc[v_s,2] \gets \text{\sc Pref}(\Mc[v_s,1] \cup \Lambda(v_s,v_s))$
\State $\Pb[v_s] = v_s$
\Else
\State $\Mc[v_s,:] \gets (\APRem,\infty)$
\EndIf
\EndIf \label{alg:haprp:sub:end}
\While {$\Qc \neq \emptyset$} \label{alg:haprp:sub:while}
\State \Comment{Get node $u$ with minimum $\Mc[u,2]$}
\State $u \gets$ {\sc ExtractMIN}($\Qc$) \label{alg:haprp:sub:min}
\If {$\Mc[u,2] < \infty$}
\State $\Vc \gets \Vc \cup \{u\}$
\For {$v \in V$ such that $(u,v) \in E$} \label{alg:haprp:sub:for2}

\State $\tupleof{\Mc,\Pb} \gets \text{\sc
Relax}((u,v),\Mc,\Pb,\Lambda)$ \label{alg:haprp:sub:relax}
\EndFor
\EndIf
\EndWhile
\State \Return $\Mc$, $\mathbf{P}$, $\Vc$
\EndProcedure \label{alg:haprp:sub:endproc}
\end{algorithmic}
\end{algorithm}
}
{
}

\ifthenelse {\boolean{BGRAPHPDF}}
{
\begin{algorithm}[tb]
\caption{{\sc Relax}}
{\bf Inputs}: an edge $(u,v)$, the tables $\Mc$ and $\Pb$ and the edge labeling function $\Lambda$ \\
{\bf Output}: the tables $\Mc$ and $\mathbf{P}$
\label{alg:haprp:relax}
\begin{algorithmic}[1]
\Procedure{{\sc Relax}}{$(u,v)$,$\Mc$,$\Pb$,$\Lambda$}
\If {$\text{\sc Pref}(\Mc[u,1] \cup \Lambda(u,v)) < \Mc[v,2]$}
\State $\Mc[v,1] \gets \Mc[u,1] \cup \Lambda(u,v)$
\State $\Mc[v,2] \gets \text{\sc Pref}(\Mc[u,1] \cup \Lambda(u,v))$
\State $\mathbf{P}[v] \gets u$
\EndIf
\State \Return $\Mc$, $\mathbf{P}$
\EndProcedure
\end{algorithmic}
\end{algorithm}
}
{
}

{\bf Correctness:} 
The correctness of the algorithm ARPP is based upon the fact that a node $v \in V$ is reachable on $G_\Ac$ if and only if $\Mc[v,2] < \infty$.
The argument for this claim is similar to the proof of correctness of Dijkstra's shortest path algorithm in \cite{CormenLRS01}.
If this algorithm returns a set of atomic propositions $L$ which removed from $\ASPEC$, then the language $\Lc(\Ac)$ is non-empty. This is immediate by the construction of the graph $G_{\Ac}$ (Def. \ref{def:graph}).


{\bf Running time:} 
The analysis of the algorithm ARPP follows closely the analysis of AAMRP in \cite{KimF12iros}.
The only difference in the time complexity is that ARPP uses \text{\sc Pref} function in order to compute preference levels of all elements in $\APRem$.
Both Min-Sum Revision and Min-Max Revision take $O(\APRem)$ since at most they compute preference levels of all elements in $\APRem$.
Hence, the running time of {\sc FindMinPath} is  $O(E (\APRem^2 \log \APRem + \log V))$.
Therefore, the running time of {\sc ARPP} is $O(V_f ( V \APRem \log \APRem + E (\APRem^2 \log \APRem + \log V))) = O(V_f E (\APRem^2 \log \APRem + \log V))$ which is polynomial in the size of the input graph.

}
{
\section{Algorithms for the Revision Problem with Preferences}
\label{sec:heuristic}

In this section, we present Algorithms for the Revision Problem with Preferences (ARPP).
It is based on the Approximation Algorithm of the Minimal Revision Problem (AAMRP) \cite{KimF12iros} which is in turn based on Dijkstra's shortest path algorithm \cite{LaValle05}.
The main difference from AAMRP is that instead of finding the minimum number of atomic propositions that must be removed from each edge on the paths of the graph $G_\Ac$,
ARPP tracks paths having atomic propositions that minimize the preference level from each edge on the paths of the graph $G_\Ac$.

Here, we present the pseudocode for ARPP.
ARPP is similar to AAMRP in \cite{KimF12iros}.
The difference from \cite{KimF12iros} is that AARP uses \text{\sc Pref} function instead of using cardinality of the set.
For Min-Sum Revision, the function \text{\sc Pref}$:\APRem \rightarrow \mathbb{R}_{\ge 0}$ is defined as following: given a set of label $R \subseteq \APRem$ and the preference function $\theta_{+}: \APRem \rightarrow \mathbb{R}_{\ge 0}$, $\text{\sc Pref}(R) = \theta_{+}(R)$.
The Min-Sum ARPP is denoted by $ARPP_{+}$.
For Min-Max Revision, the function \text{\sc Pref}$:\APRem \rightarrow \mathbb{R}_{\ge 0}$ is defined as following: given a set of label $R \subseteq \APRem$ and the preference function $\theta: \APRem \rightarrow \mathbb{R}_{\ge 0}$, $\text{\sc Pref}(R) = \max_{\rho \in R} \theta(\rho)$.
The Min-Max ARPP is denoted by $ARPP_{max}$.

The main algorithm (Alg. \ref{alg:haprp:main}) divides the problem into two tasks.
First, in line \ref{alg:main:prefix}, it finds an approximation to the minimum preference level of atomic propositions from $\APRem$ that must be removed to have a prefix path to each reachable sink (see Section \ref{sec:problem}).
Then, in line \ref{alg:main:lasso}, it repeats the process from each reachable final state to find an approximation to the minimum preference level of atomic propositions from $\APRem$ that must be removed so that a lasso path is enabled.
The combination of prefix/lasso that removes the least preferable atomic propositions is returned to the user.
Due to space limitations, we omit Algorithm 2 \text{\sc FindMinPath} and Algorithm 3 \text{\sc Relax} (for details, see \cite{KimF13icraFull}). 

\ifthenelse {\boolean{BGRAPHPDF}}
{
\begin{algorithm}[tb]
{\bf Inputs}: a graph $G_\Ac = (V,E,v_s,V_f,\APRem,\Lambda,\theta)$. \\
{\bf Outputs}: the list $L$ of symbols from $\APRem$ that must be removed from $\ASPEC$. 
\caption{ARPP}
\label{alg:haprp:main}
\begin{algorithmic}[1]
\Procedure{ARPP}{$G_\Ac$}
\State $L \gets \APRem$
\State $\Mc[:,:] \gets (\APRem,\infty)$ \Comment{Each row is set to $(\APRem,\infty)$}
\State $\Mc[v_s,:] \gets (\emptyset,0)$ \Comment{Initialize the source node}
\State $\tupleof{\Mc, \Pb, \Vc} \gets  \text{\sc FindMinPath}(G_\Ac,\Mc,0)$ \label{alg:main:prefix}
\If {$\Vc \cap V_f = \emptyset$}
$L \gets \emptyset$
\Else
\For {$v_f \in \Vc \cap V_f$} \label{alg:main:loop}
\State $L_p \gets \text{\sc GetAPFromPath}(v_s,v_f,\Mc,\Pb)$  \label{alg:main:lasso}
\State $\Mc'[:,:] \gets (\APRem,\infty)$
\State $\Mc'[v_f,:] \gets \Mc[v_f,:]$
\State $G_\Ac' \gets (V,E,v_f,\{v_f\},\APRem,L)$
\State $\tupleof{\Mc', \Pb', \Vc'} \gets  \text{\sc FindMinPath}(G'_\Ac,\Mc',1)$
\If {$v_f \in \Vc'$}
\State $L_l \gets \text{\sc GetAPFromPath}(v_f,v_f,\Mc',\Pb')$
\If {\text{\sc Pref}$(L_p \cup L_l) \leq \text{\sc Pref}(L)$} \label{alg:main:getpref}
\State $L \gets L_p \cup L_l$ 
\EndIf
\EndIf
\EndFor
\EndIf
\State \Return $L$
\EndProcedure
\end{algorithmic}
The function \text{\sc GetAPFromPath}($(v_s,v_f,\Mc,\Pb)$) returns the atomic propositions that must be removed from $\ASPEC$ in order to enable a path on $\Ac$ from a starting state $v_s$ to a final state $v_f$ given the tables $\Mc$ and $\Pb$.
\end{algorithm}
}
{
}

The analysis of the algorithm ARPP follows closely the analysis of AAMRP in \cite{KimF12iros}.
The only difference in the time complexity is that ARPP uses \text{\sc Pref} function in order to compute preference levels of all elements in $\APRem$.
Both Min-Sum Revision and Min-Max Revision take $O(\APRem)$ since at most they compute preference levels of all elements in $\APRem$.
Hence, the running time of {\sc FindMinPath} is  $O(E (\APRem^2 \log \APRem + \log V))$.
Therefore, the running time of {\sc ARPP} is $O(V_f ( V \APRem \log \APRem + E (\APRem^2 \log \APRem + \log V))) = O(V_f E (\APRem^2 \log \APRem + \log V))$ which is polynomial in the size of the input graph.

}
\section{Example and Experiments}
\label{sec:experiments}

In this section, we present an example scenario and experimental results using our prototype implementation of algorithms and brute-force search.

In the following example, we will be using LTL as a specification language.
We remark that the results presented here can be easily extended to LTL formulas by renaming repeated occurrences of atomic propositions in the specification and adding them on the transition system (for details, see \cite{LTL2BA_CPSLAB}).

The following example scenario was inspired by \cite{UlusoyEtAl2011iros, UluosoySDB2012}, and we will be using LTL as a specification language.

\ifthenelse {\boolean{BGRAPHPDF}}
{
  \begin{figure}
  \centering
  \subfloat{
    \includegraphics[width=3.5cm]{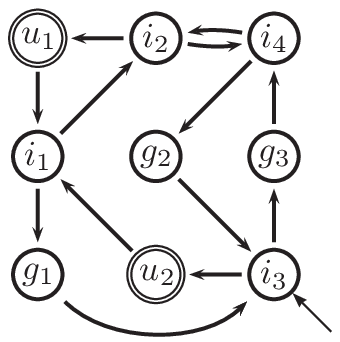}
  }
  \subfloat{
    \includegraphics[width=4.5cm]{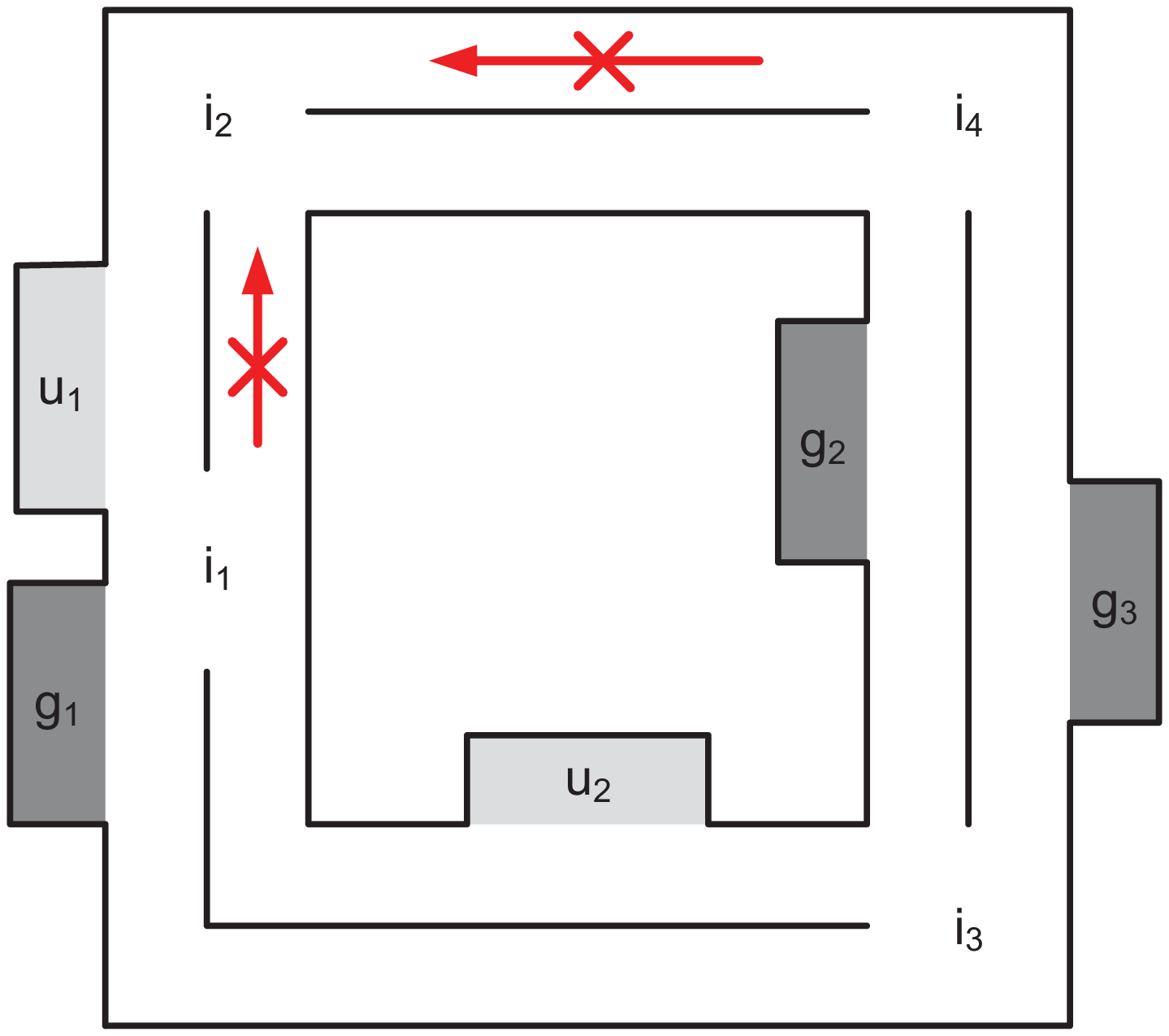}
  }
  \caption{Illustration of the simple road network environment of Example \ref{exm:single_robot_gathering}. The robot is required to drive right-side of the road.}
  \label{pic:single_robot_road}
  \end{figure}
}
{
  \begin{figure}
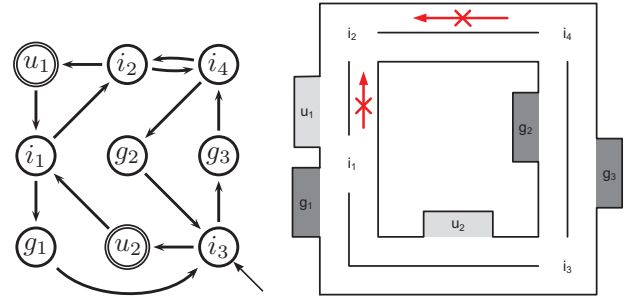

  \begin{center}
  \VCDraw{%
  \begin{VCPicture}{(0,-2.5)(4.2,2.5)}
  \FinalState[u_1]{(0,2)}{u_1} \FinalState[u_2]{(2,-2)}{u_2}
  \State[i_1]{(0,0)}{i_1}
  \State[i_2]{(2,2)}{i_2}
  \State[i_3]{(4,-2)}{i_3}
  \State[i_4]{(4,2)}{i_4}
  \State[g_1]{(0,-2)}{g_1}
  \State[g_2]{(2,0)}{g_2}
  \State[g_3]{(4,0)}{g_3}
  \ncarc[arcangle=0,nodesep=0.1cm]{->}{i_3}{u_2}
  \ncarc[arcangle=0,nodesep=0.1cm]{->}{u_2}{i_1}
  \ncarc[arcangle=0,nodesep=0.1cm]{->}{i_1}{g_1}
  \ncarc[arcangle=-45,border=0.01cm,nodesep=0.2cm]{->}{g_1}{i_3}
  \ncarc[arcangle=0,nodesep=0.1cm]{->}{i_1}{i_2}
  \ncarc[arcangle=0,nodesep=0.1cm]{->}{i_2}{u_1}
  \ncarc[arcangle=0,nodesep=0.1cm]{->}{u_1}{i_1}
  \ncarc[arcangle=-10,nodesep=0.1cm]{->}{i_2}{i_4}
  \ncarc[arcangle=-10,nodesep=0.1cm]{->}{i_4}{i_2}
  \ncarc[arcangle=0,nodesep=0.1cm]{->}{i_4}{g_2}
  \ncarc[arcangle=0,nodesep=0.1cm]{->}{g_2}{i_3}
  \ncarc[arcangle=0,nodesep=0.1cm]{->}{i_3}{g_3}
  \ncarc[arcangle=0,nodesep=0.1cm]{->}{g_3}{i_4}
  \Initial[se]{i_3}
  \end{VCPicture}}
  \end{center}
  \caption{Illustration of the simple road network environment of Example \ref{exm:single_robot_gathering}. The robot is required to drive right-side of the road.}
  \label{pic:single_robot_road_graph}
  \end{figure}
}


\begin{exmp}[Single Robot Data Gathering Task]
\label{exm:single_robot_gathering}

In this example, we use a simplified road network having three gathering locations and two upload locations with four intersections of the road. In Fig. \ref{pic:single_robot_road}, the data gather locations, which are labeled $g_1$, $g_2$, and $g_3$, are dark gray, the data upload locations, which are labeled $u_1$ and $u_2$, are light gray, and the intersections are labeled $i_1$ through $i_4$.
In order to gather data and upload the gather-data persistently, the following LTL formula may be considered:
$\phi_A$ $:=$ GF($\varphi$) $\wedge$ GF($\pi$), where $\varphi := g_1 \vee g_2 \vee g_3$ and $\pi := u_1 \vee u_2$.
The following formula can make the robot move from gather locations to upload locations after gathering data:
$\phi_G$ $:=$ G($\varphi \rightarrow$ X($\neg\varphi \Un \pi$)).
In order for the robot to move to gather location after uploading, the following formula is needed:
$\phi_U$ $:=$ G($\pi \rightarrow$ X($\neg\pi \Un \varphi$)).

Let us consider that some parts of road are not recommended to drive from gather locations, such as from $i_4$ to $i_2$ and from $i_1$ to $i_2$.
We can describe those constraints as follows:
$\psi_1$ $:=$ G($g_1$ $\rightarrow$ $\neg$($i_4$ $\wedge$ X$i_2$)$\Un$$u_1$) and
$\psi_2$ $:=$ G($g_2$ $\rightarrow$ $\neg$($i_1$ $\wedge$ X$i_2$)$\Un$$u_2$).
If the gathering task should have an order such as $g_3$, $g_1$, $g_2$, $g_3$, $g_1$, $g_2$, $\ldots$, then the following formula could be considered:
$\phi_O$ := 
(($\neg$$g_1$ $\wedge$ $\neg$$g_2$)$\Un$$g_3$) $\wedge$
G($g_3$ $\rightarrow$ X(($\neg$$g_2$ $\wedge$ $\neg$$g_3$)$\Un$$g_1$)) $\wedge$
G($g_1$ $\rightarrow$ X(($\neg$$g_1$ $\wedge$ $\neg$$g_3$)$\Un$$g_2$)) $\wedge$
G($g_2$ $\rightarrow$ X(($\neg$$g_1$ $\wedge$ $\neg$$g_2$)$\Un$$g_3$)).
Now, we can informally describe the mission. The mission is ``Always gather data from $g_3$, $g_1$, $g_2$ in this order and upload the collected data to $u_1$ and $u_2$. Once data gathering is finished, do not visit gather locations until the data is uploaded. Once uploading is finished, do not visit upload locations until gathering data. You should always avoid the road from $i_4$ to $i_2$ when you head to $u_1$ from $g_1$ and from $i_1$ to $i_2$ when you head to $u_2$ from $g_2$''. The following formula represents this mission:
\begin{center}
$\phi_{single}$ := $\phi_O \wedge \phi_G \wedge \phi_U \wedge \psi_1 \wedge \psi_2 \wedge$ GF($\pi$).
\end{center}


Assume that initially, the robot is in $i_3$ and final nodes are $u_1$ and $u_2$.
When we made a cross product with the road and the specification, we could get 36824 states, 350114 transitions and 100 final states.
Not removing some atomic propositions, the specification was not satisfiable.

We tested two different preference levels.
For clarity in presentation,
we omit for presenting preference levels on each transition since we set for all the occurances of the same symbols the same preference level, we abuse notation and write $\theta(\pi)$ instead of $\theta(\pi,(s_i,s_j))$. However, the revision is for specification transitions.
First, the preference level of the symbols are as follows: for $g_1$, $g_2$, $g_3$, $u_1$, $u_2$, $i_1$, $i_2$, $i_3$, $i_4$, the preference levels are 3, 4, 5, 20, 20, 1, 1, 1, 1, respectively, and for $\neg g_1$, $\neg g_2$, $\neg g_3$, $\neg u_1$, $\neg u_2$, $\neg i_1$, $\neg i_2$, $\neg i_3$, $\neg i_4$, the preference levels are 3, 4, 5, 20, 20, 1, 1, 1, 1, respectively.  ARPP for Min-Sum Revision took 210.979 seconds, and suggested removing $\neg g_1$ and $\neg i_4$. The total returned preference was 4 since $\theta(\neg g_1) = 3$ and $\theta(\neg i_4) = 1$.
The sequence of the locations suggested by ARPP is 
$i_3$$g_3$$i_2$$u_1$$($$i_1$$g_1$$i_3$$u_2$$i_1$$i_2$$i_4$$g_2$$i_3$$u_2$$i_1$$g_1$$i_3$$g_3$$i_4$$i_2$$u_1$$)^+$.
We can check that $\neg g_1$ is from $G(g_2 \rightarrow X((\neg g_1 \land \neg g_2)\Un g_3))$ of the formula $\phi_O$
and from $\neg \varphi = \neg(g_1 \vee g_2 \vee g_3)$ of the formula $\phi_G = G(\varphi \rightarrow (\neg \varphi \Un \pi))$,
and $\neg i_4$ is from $G(g_1 \rightarrow \neg(i_4 \wedge X i_2) \Un u_1)$ of the formula $\psi_1$.
AARP for Min-Max Revision took 239 seconds, and returned $g_1$, $\neg g_1$, $\neg i_1$, and $\neg i_4$.
The maximum returned preference was 3 since $\theta(g_1) = 3$  and $\theta(\neg g_1) = 3$.


In the second case, the preference level of the positive atomic propositions are same as the first test, and the preference level of the negative atomic propositions are as follow: for $\neg g_1$, $\neg g_2$, $\neg g_3$, $\neg u_1$, $\neg u_2$, $\neg i_1$, $\neg i_2$, $\neg i_3$, $\neg i_4$, the preference levels are 3, 4, 5, 20, 20, 10, 10, 10, 10, respectively.  In this case, ARPP for Min-Sum Revision took 207.885 seconds, and suggested removing $g_3$.
The total returned preference was 5 since $\theta(g_3) = 5$.
The sequence of the locations suggested by ARPP is $i_3 g_3 i_4 i_2 u_1 (i_1 g_1 i_3 u_2 i_1 i_2 i_4 g_2 i_3 u_2 i_1 i_2 u_1)^+$. We can check that $g_3$ is from $G(g_3 \rightarrow X((\neg g_2 \wedge \neg g_3)\Un g1))$ of the formula $\phi_O$ and from $\varphi = (g_1 \vee g_2 \vee g_3)$ of the formula $\phi_U = G(\phi \rightarrow X(\neg \phi \Un \varphi)$.
ARPP for Min-Max Revision took 214.322 seconds, and returned $g_1$ and $\neg g_1$.
The maximum preference was 3 since $\theta(g_1) = 3$ and $\theta(\neg g_1)= 3$.
\exmend
\end{exmp}

\ifthenelse {\boolean{TECHREP}}
{
\begin{table*}
\begin{center}
\begin{tabular}{|c|c|c|c|c|c|c|c|c|c|c|c|}
\hline
Nodes & \multicolumn{4}{c|} {Brute-Force} & \multicolumn{4}{c|} {Min-Sum Revision} & \multicolumn{3}{c|} {RATIO} \\
\hline
      &    min &   avg &   max &   succ &   min &   avg &   max &   succ &   min &   avg &   max \\
\hline
9     & 0.033  & 0.0921 & 0.945 & 200/200 & 0.019 & 0.183 & 0.874 & 200/200 & 1 & 1 & 1 \\
\hline
100   & 0.065  & 0.3707 & 3.997 & 200/200 & 0.065 & 0.1598 & 2.66 & 200/200 & 1 & 1.003 & 1.619 \\
\hline
196   & 0.278  & 303.55 & 11974 & 199/200 & 0.137 & 0.4927 & 12.057 & 200/200 & 1 & 1.0014 & 1.1475 \\
\hline
\end{tabular}
\end{center}
\caption{Numerical Experiments: Number of nodes versus the results of Brute-Force Search Algorithm and ARPP for Min-Sum Revision.
Under the Brute-Force and Min-Sum Revision columns the numbers indicate computation times in $\sec$.
RATIO indicates the experimentally observed approximation ratio to the optimal solution.}
\label{experiment_result_dag1}
\end{table*}

\begin{table*}
\begin{center}
\begin{tabular}{|c|c|c|c|c|c|c|c|c|c|c|c|c|c|c|}
\hline
Nodes & \multicolumn{4}{c|} {Min-Sum Revision ($ARPP_{+}$)} & \multicolumn{4}{c|} {Min-Max Revision ($ARPP_{max}$)} & \multicolumn{3}{c|} {RATIO1} & \multicolumn{3}{c|} {RATIO2} \\
\hline
      &     min &    avg &    max &    succ &    min &    avg &    max &    succ &    min &     avg &    max &    min &    avg &    max \\
\hline
9     &  0.019  &  0.183 &  0.874 & 200/200 &  0.02  & 0.0508 &  0.66  & 200/200 & 1      & 1.2677  & 3.4    &   1    & 1.0007 & 1.1428 \\
\hline
100   &  0.065  & 0.1598 &  2.66  & 200/200 & 0.061  & 0.1258 & 0.471  & 200/200 & 1      & 1.441   & 5.97   &   1    & 1.0264 & 1.3928 \\
\hline
196   &  0.137  & 0.4927 & 12.057 & 200/200 & 0.139  & 0.29824 & 0.74  & 200/200 & 1      & 1.4876  & 5.634  &   1    & 1.0389 & 2.1904 \\
\hline
\end{tabular}
\end{center}
\caption{Numerical Experiments: For each graph $G_\Ac$, Number of nodes versus the results of ARPP for Min-Sum Revision ($ARPP_{+}$) and ARPP for Min-Max Revision ($ARPP_{max}$).
Under the Min-Sum Revision and Min-Max Revision columns the numbers indicate computation times in $\sec$.
RATIO1 indicates $\sum(\theta(ARRP_{max}(G_\Ac)))/\sum(\theta(ARPP_{+}(G_\Ac)))$.
RATIO2 indicates $max(\theta(ARRP_{+}(G_\Ac))/max(\theta(ARPP_{max}(G_\Ac)))$.
}
\label{experiment_result_dag2}
\end{table*}

Now, we present experimental results. The propotype implementation is written in Python.

For the experiments, we utilized the ASU super computing center which consists of clusters of Dual 4-core processors, 16 GB Intel(R) Xeon(R) CPU X5355 @2.66 Ghz. 
Our implementation does not utilize the parallel architecture. 
The clusters were used to run the many different test cases in parallel on a single core. 
The operating system is CentOS release 5.9.

In order to assess the experimental approximation ratio of the heuristic (Min-Sum Revision), we compared the solutions returned by the heuristic with the Brute-force search.
The Brute-force search is guaranteed to return a minimal solution to the Min-Sum Revision problem.

We performed a large number of experimental comparisons on random benchmark instances of various sizes.
Each test case consisted of two randomly generated DAGs which represented an environment and a specification.
Both graphs have self-loops on their leaf nodes so that a feasible lasso path can be found.
The number of atomic propositions in each instance was equal to four times the number of nodes in each acyclic graph.
For example, in the benchmark where the graph had 9 nodes, each DAG had 3 nodes, and the number of atomic propositions was 12.
The final nodes are chosen randomly and they represent 5\%-40\% of the nodes.
The number of edges in most instances were 2-3 times more than the number of nodes.

Table \ref{experiment_result_dag1} compares the results of the Brute-Force Search Algorithm with the results of ARPP for Min-Sum Revision on test cases of different sizes (total number of nodes). For each graph size, we performed 200 tests and we report minimum, average, and maximum computation times in $\sec$. Both algorithms were able to finish the computation and return a minimal revision for instances having 9 nodes and 100 nodes. However, for instances having 196 nodes, the Brute-Force Search Algorithm had one failed instance which exceeded the 2 hrs window limit. In the large problem instances, ARPP for Min-Sum Revision achieved a 600 time speed-up on the average running time.

%
In Table \ref{experiment_result_dag2}, we present two ratios. RATIO1 captures the ratios between the sum of preference levels of the set returned by $ARPP_{max}$ over the sum of preference levels of the set returned by $ARPP_{+}$. On the other hand,  RATIO2 captures the ratios between the max of preference levels of the set returned by $ARPP_{+}$ over the max of preference levels of the set returned by $ARPP_{max}$. If the $ARPP_{+}$ was always returning the optimal solution, then RATIO1 should always be greater than 1. We observe on the random graph instances that the result also holds for this particular class of random graphs. Moreover, there were graph instances where $ARPP_{+}$ returned much smaller total preference sum then $ARPP_{max}$. Importantly, when received the results for RATIO2, we observe that there exist graph instances where $ARPP_{max}$ returned a revision set with maximum much less then the maximum preference in the set returned by $ARPP_{+}$. Thus, depending on the user application it could be desirable to utilize either revision criterion.

\begin{table}
\begin{center}
\begin{tabular}{|c|c|c|c|c|c|}
\hline
Nodes & Min-Sum  & Min-Max       & \multicolumn{3}{c|}{RATIO}   \\
\hline
      & avg         & avg           & min   & avg    & max \\
\hline
9     & 1.305       & 1.785         & 0.66  & 1.423  & 5   \\
\hline
100   & 1.95        & 3.215         & 1     & 1.8056 & 6   \\
\hline
196   & 2.305       & 3.84          & 1     & 1.7793 & 8   \\
\hline
\end{tabular}
\end{center}
\caption{Numerical Experiments: Number of nodes versus the results of ARPP for Min-Sum Revision ($ARPP_{+}$) and ARPP for Min-Max Revision ($ARPP_{max}$).}
\label{experiment_result_dag3}
\end{table}

Table \ref{experiment_result_dag3} shows the comparison between the number of atomic propositions of the set returned from ARPP for Min-Sum Revision ($ARPP_{+}$) and the number of atomic propositions of the set returned from ARPP for Min-Max Revision ($ARPP_{max}$).
The columns under the avg columns of Min-Sum and Min-Max indicate the average number of atomic propositions of the set returned from $ARPP_{+}$ and $ARPP_{max}$ for graph instances having 9 nodes, 100 nodes, and 196 nodes.
The RATIO captures the ratios between the number of atomic propositions of the set returned by $ARPP_{max}$ over the number of atomic propositions of the set returned by $ARPP_{+}$.
Even though Min-Sum Revision and Min-Max Revision do not count the number of atomic propositions while relaxing, this result shows readers how many atomic propositions each algorithm returns.
From the fact that the avg of the RATIO for all random graph instances is greater than 1, we observe that the set returned from Min-Max Revision in general has more number of atomic propositions than the set returned from Min-Sum Revision.


}
{
\begin{table*}
\begin{center}
\begin{tabular}{|c|c|c|c|c|c|c|c|c|c|c|c|c|}
\hline
Nodes & \multicolumn{4}{c|} {Brute-Force} & \multicolumn{5}{c|} {Min-Sum Revision ($ARPP_{+}$)}          & \multicolumn{3}{c|} {RATIO} \\
\hline
      &    min &   avg  &   max &   succ  &   min &   avg  &   max  &   succ  & avg \# nodes &  min &   avg  &   max       \\
\hline
9     & 0.033  & 0.0921 & 0.945 & 200/200 & 0.019 & 0.183  & 0.874  & 200/200 &  1.305      & 1    & 1      & 1           \\
\hline
100   & 0.065  & 0.3707 & 3.997 & 200/200 & 0.065 & 0.1598 & 2.66   & 200/200 &  1.95       & 1    & 1.003  & 1.619       \\
\hline
196   & 0.278  & 303.55 & 11974 & 199/200 & 0.137 & 0.4927 & 12.057 & 200/200 &  2.305      & 1    & 1.0014 & 1.1475      \\
\hline
\end{tabular}
\end{center}
\caption{Numerical Experiments: Number of nodes versus the results of Brute-Force Search Algorithm and ARPP for Min-Sum Revision.
Under the Brute-Force and Min-Sum Revision columns the numbers indicate computation times in $\sec$.
RATIO indicates the experimentally observed approximation ratio to the optimal solution.}
\label{experiment_result_dag1}
\end{table*}
Now, we present some experimental results. The propotype implementation is written in Python.
For the experiments, we utilized the ASU super computing center which consists of clusters of Dual 4-core processors, 16 GB Intel(R) Xeon(R) CPU X5355 @2.66 Ghz. 
The operating system is CentOS release 5.9.
The clusters were used to run each test case on each single core in parallel.

In order to assess the experimental approximation ratio of the heuristic (Min-Sum Revision), we compared the solutions returned by the heuristic with Brute-force search algorithm.
The Brute-force search is guaranteed to return a minimal solution to the Min-Sum Revision problem.
We omit the explanation of the each test case and full experiment results (for detail, see \cite{KimF13icraFull}).


Table \ref{experiment_result_dag1} compares the results of the Brute-Force Search Algorithm with the results of ARPP for Min-Sum Revision ($ARPP_{+}$) on test cases of different sizes (total number of nodes). For each graph size, we performed 200 tests and we report minimum, average, and maximum computation times in $\sec$.
The ``avg \# nodes'' column shows the average number of nodes returned from $ARPP_{+}$.
Both algorithms were able to finish the computation and return a minimal revision for instances having 9 nodes and 100 nodes.
However, for instances having 196 nodes, the Brute-Force Search Algorithm had one failed instance which exceeded the 2 hrs window limit.
In the large problem instances, ARPP for Min-Sum Revision achieved a 600 time speed-up on the average running time.

Table \ref{experiment_result_dag2} shows the results of ARPP for Min-Max Revision ($ARPP_{max}$) on test cases of different sizes (total number of nodes). This test results also used same test cases as the ones for Table \ref{experiment_result_dag1}. We report minimum, average, and maximum computation times in $\sec$. The ``avg \# nodes'' shows the average number of nodes returned from $ARPP_{max}$.

\begin{table}[h!]
\begin{center}
\begin{tabular}{|c|c|c|c|c|c|}
\hline
Nodes & \multicolumn{5}{c|} {Min-Max Revision ($ARPP_{max}$)} \\
\hline
      &  min    &    avg  &    max &    succ & avg \# nodes   \\
\hline
9     &  0.02   & 0.0508  &  0.66  & 200/200 & 1.785          \\
\hline
100   &  0.061  & 0.1258  & 0.471  & 200/200 & 3.215          \\
\hline
196   &  0.139  & 0.29824 & 0.74   & 200/200 & 3.84           \\
\hline
\end{tabular}
\end{center}
\caption{Numerical Experiments: For each graph $G_\Ac$, Number of nodes versus the result of ARPP for Min-Max Revision ($ARPP_{max}$).
Under the min, avg, max columns the numbers indicate computation times in $\sec$.
}
\label{experiment_result_dag2}
\end{table}

}

\section{Conclusions}

This paper discusses the problem of specification revision with user preferences.
We have demonstrated that adding preference levels to the goals in the specification can render the revision problem easier to solve under the appropriate cost function.
We view the automatic debugging and specification revision problems as foundational for formal methods to receive wider adoption in the robotics community and beyond.
With the current paper and the predecessors \cite{Fainekos11icra,KimF12iros,KimFS12icra,KimF13icra}, we have studied the theoretical foundations of different versions of the problem.
Our algorithms and tools can be used as add-ons to control synthesis methods developed by our and other groups \cite{FainekosGKGP09automatica,LaViersEtAl2011iccps,UlusoyEtAl2011iros,BobadillaEtAlRSS11,WolffTM13iros}.
Our goal for the future is to incorporate all the specification revision methods in a comprehensive user-friendly tool that can run on different platforms.



\section*{Acknowledgments}
The authors would like to thank the anonymous reviewers for their detailed comments.

\bibliographystyle{IEEEtran}
\bibliography{robotics_bibrefs,fainekos_papers}

\end{document}